\documentclass[8.5pt,twoside,twocolumn]{article}
\usepackage{amsmath,amssymb}
\usepackage{graphicx}

\usepackage[super,sort&compress,comma]{natbib} 

\begin{document}
	\twocolumn[
	\begin{@twocolumnfalse}
		\noindent\LARGE{\textbf{Scale dependence of mechanics of active gels with increasing motor concentration}}
		\vspace{0.6cm}
		
		\noindent\large{\textbf{Adar Sonn-Segev,\textit{$^{a}$} Anne Bernheim-Groswasser, \textit{$^{b}$} and Yael Roichman$^{\ast}$\textit{$^{a}$}}}\vspace{0.5cm}
		\noindent \normalsize{Actin is a protein that plays an essential role in maintaining the cell’s mechanical integrity. In response to strong external stresses, it can assemble into large bundles, but grows into a fine branched network to induce cell motion. In some cases, the self-organization of actin fibers and networks involves the action of bipolar filaments of the molecular motor myosin. Such self-organization processes mediated by large myosin bipolar filaments were studied extensively in-vitro.  Here we create active gels, comprised of single actin filaments and small myosin bipolar filaments. The active steady state in these gels persists long enough to enable the characterization of their mechanical properties using one and two point microrheology. We study the effect of myosin concentration on the mechanical properties of this model system for active matter, for two different motor assembly sizes. Contrary to previous studies of networks with large motor assemblies, we find that the fluctuations of tracer particles embedded in the network decrease in amplitude as motor concentration increases. Nonetheless, we show that myosin motors stiffen the actin networks, in accord with bulk rheology measurements of networks containing larger motor assemblies.  This implies that such stiffening is of universal nature and may be relevant to a wider range of cytoskeleton-based structures. 
		}
		\vspace{0.5cm}
	\end{@twocolumnfalse}
]

\footnotetext{\textit{$^{a}$~Raymond \& Beverly Sackler School of Chemistry, Tel Aviv University, Tel Aviv 6997801, Israel; E-mail: roichman@tau.ac.il}}
\footnotetext{\textit{$^{b}$~Department of Chemical Engineering, Ilse Kats Institute for Nanoscale Science and Technology, Ben Gurion University of the Negev, Beer-Sheva 84105, Israel}}

\section{Introduction}
Active gels inspired from the cell's skeleton are a paradigmatic model used to understand the complex mechanisms governing key cellular processes such as motion. These active gels are also attractive as model systems to study active matter and non-equilibrium statistical mechanics. A well studied example of such materials are networks constructed from the structural protein actin and its associated molecular motors myosin II\cite{Koenderink2009,Mizuno2007,Toyota2011,Gardel2011,Koehler2012,Liverpool2001,Humphrey2002,Ziebert2008,Backouche2006,Bendix2008,Ideses2013,Kohler2011a}.
Both actin and myosin~II are key components in cell motility and muscle contraction. Myosin motor domains (heads) generate active motion by hydrolysis of chemical fuel in the form of adenosine triphosphate (ATP). The hydrolysis process promotes a configurational change of the myosin, which results in stepwise walking along the actin filament. One important and unique feature of myosin~II is its ability to form multimeric bipolar structures containing between tens to hundreds of myosin molecules\cite{Reisler1980,Sinard1989,Ideses2013}. The number of myosin heads within such a bipolar filament can be tuned by the salt concentration in the self-organization buffer \cite{Reisler1980,Sinard1989,Ideses2013}. This structure allows the myosin bipolar filaments to connect multiple actin filaments simultaneously and move them relative to each other \cite{Robert1977,HAYASHI1975}. In an entangled actin network this  motion generates flow \cite{Liverpool2001,Humphrey2002,Ziebert2008}, whereas in a crosslinked network it generates internal contractile forces \cite{Humphrey2002,Backouche2006,Koenderink2009,Bendix2008,Ideses2013,Kohler2011a,Koehler2012,Lenz2014}. Even without myosin, the mechanical properties of actin networks depend on details such as their filament length \cite{Janmey1994} and  crosslinking density \cite{Wagner2006}. For example, the type of crosslinker determines the structural organization of the actin into a mesh of single or bundled actin filaments \cite{Lieleg2010}, and in turn affects the gel's mechanical properties \cite{Gardel2008,Kasza2009}. The ability of these proteins to self-organize into structures of different characteristic scales is of fundamental importance to their various functions, hence their remarkable diverse dynamical behaviors including reorganization\cite{Silva2011,Ideses2013,Backouche2006}, structural evolution \cite{Ideses2013,Backouche2006}, global compression\cite{Koehler2012,Ideses2013,Bendix2008,Linsmeier2016}, and rupture\cite{Alvarado2013,Ideses2013}.

\begin{figure}[h!]
\centering
\includegraphics[scale=0.35]{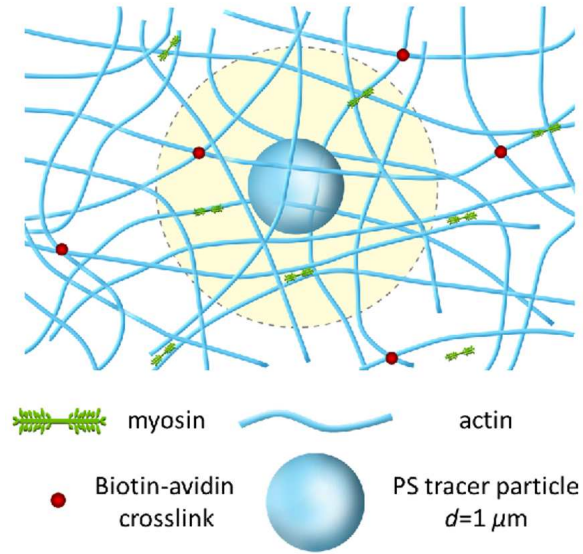}
\caption{ Schematic illustration of the active gels and its components. Myosin bipolar filaments are embedded within a network of actin filament crosslinked by biotin-neutravidin bond. Tracer particles are used as a probe for the network's mechanical properties. }
\label{fig:gel}
\end{figure}

In order to observe the rich dynamics of active actomyosin networks directly, commonly studied networks involve large motor assemblies (hundreds of myosin heads per aggregate) and actin bundles which can be observed in fluorescent microscopy \cite{Kohler2011,Murrell2012,Alvarado2013,Ideses2013}. While enabling direct characterization of the networks' dynamics, these structures resemble only a subset of cytoskeleton networks found in living cells. It should be noted, that the kinetic properties and filament size  of muscle myosin II, used in these studies, are different from the nonmuscle myosin II found in cells\cite{Alcala2016,Billington2013}. Nonetheless, these studies provided crucial insight into the underlying physical processes governing such active matter systems. For example, it was shown that large myosin bipolar filaments can exert significant forces on the actin network causing their structure to evolve \cite{Koenderink2009,Koehler2012,Backouche2006,Ideses2013} rapidly.  Due to the continuous evolution of these gels, which ends with catastrophic collapse or rupture, the investigation of their motor-induced mechanical properties is challenging. One reason for the catastrophic end of such gels, in-vitro, may be the absence of a stress release mechanism, such as the continuous process of polymerization and depolymarization of actin in cells, i.e. actin turnover\cite{Carlier2017}. Previous studies of networks with large myosin filaments  concentrated, therefore, on transient properties \cite{Silva2014}. Only a few experiments were performed in steady state conditions concentrating on the fluctuation-dissipation relation breakdown due to active fluctuations at a given motor concentration and low ATP concentrations \cite{Mizuno2007,Mizuno2008}. Here, we present a systematic study of long-lived active gels containing different motor concentrations. This is achieved by creating gels with fine structural features, i.e, single actin filaments locally crosslinked by biotin-neutravidin bonds (but without bundling agents) and small motor protein aggregates of $\approx 20 -30 $ dimers (Fig.~\ref{fig:gel}). These conditions allow us to characterize the effect of motor protein concentration on the mechanical properties of the gels. We characterize their structural and mechanical properties using microrheology of tracer particles, and find that the mechanical effect of our small myosin bipolar filaments on an actin network is similar to the effect of large bipolar filaments which were reported previously, namely, the motors induce network stiffening. However, the network fluctuations in our gels decrease with motor concentration in contrast to their increase in networks containing large myosin bipolar filaments \cite{Kohler2011, Silva2014}.

\subsection{Experimental}
Active actin-myosin networks were reconstituted \textit{in vitro} by polymerizing G-actin in the presence of biotin-neutravidin crosslinkers and myosin II bipolar filaments.
G-actin was purified from rabbit skeletal muscle acetone powder \cite{Spudich1971}, with a gel filtration step, stored on ice in G-buffer (5 mM Tris HCl, 0.1 mM CaCl$_2$, 0.2 mM ATP, 1 mM DTT, 0.01\% NaN$_3$, $p$H 7.8) and used within two weeks. Purification of rabbit skeletal muscle myosin II  was done according to standard protocols \cite{Margossian1982}. Full length His-tag (N-terminal) mouse $\alpha1\beta2$ capping protein (kind gift of Pekka Lappalainen) was expressed and purified from BL21(DE3) Escherichia coli. Briefly, the E.coli cells were sonicated and centrifuged. The clarified supernatant was run through a nickel affinity column followed by anion exchange and gel filtration. 
The concentration of the G-actin,  myosin II, and capping protein was determined by absorbance measurement using a UV/Visible spectrophotometer (Ultraspec 2100 pro, Pharmacia) in a cuvette with a 1 cm path length and the extinction coefficients:  G-actin - $\epsilon_{290}=26,460$ M$^{-1}$cm$^{-1}$ \cite{Houk1974},  two-headed myosin II - $\epsilon_{280}=268,800$ M$^{-1}$cm$^{-1}$, and capping protein $\epsilon_{280} = 41430$ M$^{-1}$cm$^{-1}$ (calculated).
Myosin II dimers (stock concentration of $20~\mu$M) were stored at -80$^\text{o}$C in high salt buffer (0.5 KCl, $35\%$ sucrose) that preserves them in a dimeric form. Biotinylated actin (Cytoskeleton, Inc.) and neutravidin (Invirogen, used as a crosslinker) were mixed in G-Buffer with unlabeled actin and left over ice for at least one hour. The mixing ratios were unlabeled: biotinylated: neutravidin = 5000:5:2 giving a total actin concentration of 24 $\mu$M. The average distance between biotin/neutravidin crosslinks is estimated by $l_c=[\text{Actin}]/370*[\text{biotin}]\approx 3~\mu$m assuming 370 actin monomers per one micron actin filament length.
Small bipolar filaments are formed by diluting the myosin solution ($0.5$ MKCl) with G-buffer to the desired KCl concentration, $0.13$~M or $0.1$~M, corresponding to motor bipolar filaments, each composed of $N_{\text{myo}}=19\pm4$ or $32\pm6$ myosin dimers, respectively (Fig. S1 and \cite{Siton-Mendelson2014,Ideses2013}). KCl concentration may also affect the binding affinity and associated kinetic constant of myosin to actin. However, in the range of KCl concentration used here, these effects are minor \cite{Bechet1985}.
Polystyrene colloids with a radius of $0.55~\mu$m (Invitrogen, Lot \#742530) were incubated with a 10 mg/ml BSA solution to prevent non-specific binding of proteins to the beads' surface \cite{Valentine2001} prior to mixing with the actin solution. We set the average filament length to be $\approx 13~\mu$m by addition of 5~nM capping proteins.    

Active network formation was initiated by adding the actin solution, myosin bipolar filaments in various concentrations, capping protein, and tracer particles ($10^9$ beads/ml) to the motility buffer (10 mM HEPES, 1 mM MgCl$_2$, 0.1 mM MgATP, 0.5 mg/mL creatine kinase, 5 mM creatine phosphate, 0.2  mM EGTA, and 0.1 or 0.13 M KCl). Creatine kinase and phosphate were used as an ATP regenerating system. Immediately after polymerization the sample was infused into a glass cell, 150 $\mu$m high, and sealed with grease. The glass surfaces were coated with methoxy-terminated polyethylene glycol (PEG) to prevent binding of the proteins to the glass. The passivation procedure included cleaning the glass coverslips using Piranha, sonication in deionized water, and incubation over night with 3-mercaptopropylilane (Aldrich) in methanol/acetic acid solution (''Silane'' solution). These coverslips were kept in the silane solution for up to a week. At the day of the experiments, the glass surfaces were finally passivated with methoxy-PEG-maleimide (mPEG-mal Mw=5000 g/mol (Nanocs)) in phosphate buffered saline for at least two hours at room temperature. Shortly after cell loading, samples were fluorescently imaged at $\lambda=605$ nm with a 40$ \times $ air objective.
Each sample was monitored for approximately 160 min, in which short movies were taken every fifteen minutes, starting from $\sim 5$ min after polymerization. To avoid wall effects, imaging was done at a plane distanced at least 80 $\mu$m from the cell walls. Particle motion was recorded using a CMOS video camera (Gazelle, Point Gray) at a frame rate of 70~Hz and was tracked with accuracy of at least 13 nm using conventional video microscopy algorithms \cite{Crocker1996}. Experiments were repeated several times. Analysis of two full sets of experiments are presented here, in which each data point in steady state is averaged over 8 measurements.

\section{Results}

\subsection{Actin networks in active steady state}

In our experimental system, the actomyosin networks are kept mechanically stable by the addition of passive, chemical crosslinkers (biotin-neutravidin bonds). Our purpose is to generate active gels with a varying degree of activity maintaining steady state dynamics for long durations in order to collect sufficient statistics on motor activity. Towards this end, we keep the network structural features fine, refraining from bundling the actin filaments by the choice of a suitable crosslinker \cite{Lieleg2010}. We use small bipolar filaments of motor proteins containing only tens (Fig. S1) of two-headed myosin molecules, and provide a low but constant concentration of ATP (0.1 mM). In this ATP concentration the motor bipolar filaments stay connected to the actin filaments for statistically longer durations. The myosin bipolar filaments act both as crosslinkers and as active, stress inducing, elements. The pinching forces applied by these small motor filaments have a limited range due to the relatively small number of heads in each filament and the shortage in ATP. This limitation ensures that, while the motors generate active forces in the network, there is limited large-scale reorganization or deformation of the network, thereby reaching steady state dynamics. We use embedded tracer particles larger than the network's mesh size so that their motion will reflect the network's bulk mechanics. By design \cite{Schmidt1989} we work with actin concentration of 24 $\mu$M, resulting in an actin network with a mesh size of $\approx 0.3~\mu$m, the average distance between the passive chemical crosslinks in our system is on the order of 3~$\mu$m, the tracer particle size is 1~$\mu$m in diameter, and the average distance between myosin bipolar filaments embedded within the network ranges from 1.3 to 50 $\mu$m. A schematic illustration of our experimental model system is shown in Fig.~\ref{fig:gel}.
Our experiments consist of polymerization of the actin network in the presence of myosin bipolar filaments and passivated colloidal tracer particles.  Already at the first measurement occurring several minutes after the initiation of actin polymerization we find subdiffusive motion of the tracer particles (Fig. S2), indicating that the transition to a gel took place. Each gel is monitored for approximately three hours in which short movies are taken every fifteen minutes, Figure S3 (and movie S1) and Figure S4 (and movie S2) are taken one and two hours after mixing, respectively. We set out to characterize the effect of myosin concentration and bipolar filament size on the mechanical properties of the actomyosin networks, i.e., the effect of activity level. To this end, we performed two series of experiments, one with bipolar filaments containing $N_{\text{myo}}=19\pm4$ myosin molecules (i.e., double headed)  and the other with bipolar filaments containing $N_{\text{myo}}=32\pm6$ myosin molecules (Fig.~S1). We note that the exact number of motor heads may change a little with motor concentration. Myosin bipolar filament size is controlled by KCl concentration, as described in the methods section. In addition, it may also depend on myosin concentration. 
To verify that the effect of myosin concentration is subdominant we prepared networks in two ways: polymerizing actin in the presence of myosin dimers, or with preassembled myosin bipolar filaments. In both cases the final networks exhibited similar self-organization dynamics and structure. We repeated this for different KCl values and saw the same corresponding patterns. Based on these evidences we conclude that the dominating factor controlling bipolar filaments size is the salt concentration. It should be noted though, that we cannot exclude the possibility that there might be some changes (decrease) in the size of the bipolar filaments when myosin concentration is significantly reduced. Nevertheless, the fact that the bipolar filaments are still processive even at motor concentrations as low as $0.04~\mu$M, namely, they still generate tension in the network (demonstrated below), suggests that myosin aggregation still occurs at these low myosin concentration. We conclude, therefore that if changes in the size of the bipolar filaments occur, they have a relatively minor effect on the motor ability to attach to and apply forces on the network.  
The two motor bipolar filaments sizes were chosen specifically to ensure their processivity, while in the same time allowing the networks to reach a steady state. In each series of experiments we varied the myosin concentration from [Myosin]/[Actin]=0 to [Myosin]/[Actin]=0.02 ($N_{\text{myo}}=19$) and [Myosin]/[Actin]=0.01 ($N_{\text{myo}}=32$), corresponding to a ratio of myosin bipolar filaments to actin filaments $0<NF_{\text{myo}}/NF_{\text{actin}}<4$  and $0<NF_{\text{myo}}/NF_{\text{actin}}<1.5$, respectively ($NF_{\text{actin}}$ is calculated by considering that there are 370 actin monomers in a 1 $\mu$m actin filament).  The tracer particle trajectories are then extracted using conventional video microscopy techniques \cite{Crocker1996}. 
\begin{figure} [t]
\centering
{\includegraphics[scale=0.335]{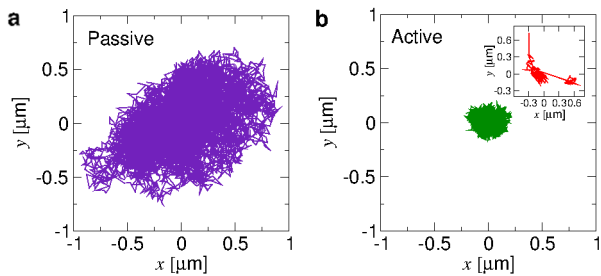}}
\caption{Typical trajectory of a 1~$\mu$m polystyrene particle in the crosslinked actin network: \textbf{a} Passive network, [Myosin]/[Actin]=0 and \textbf{b} active network, [Myosin]/[Actin]=0.02. The concentration of actin monomers is 24~$\mu$M. Motor bipolar filaments are constructed with $N_{\text{myo}}=19$ myosin molecules. The two trajectories were taken at 70~Hz for about 140~s. An example of a trajectory of a particle that experiences large steps is plotted in the inset of \textbf{b}.} \label{fig:traj}
\end{figure}
In Fig.~\ref{fig:traj} a typical trajectory of a tracer particle (140~s long) in an actin gel containing no myosin motors (passive gel), is compared to a trajectory of a bead in an active gel ([Myosin]/[Actin]=0.02) of the same duration. Both trajectories show that the tracer particle undergoes diffusive motion, however the particle embedded in the active gel seems more confined. This result is counter-intuitive, since we expect motor activity to enhance the gel's fluctuations and therefore the tracer particles' motion, as reported previously \cite{Toyota2011,LeGoff2002}. However, motor proteins are known to stiffen actin gels due to two effects: they act as additional crosslinkers and they apply tension on the actin filaments \cite{Mizuno2007}. Both processes reduce the entropy of the gel network causing it to stiffen. Therefore, we conclude that in our experiments, the added active fluctuations from the motor's activity have less effect on the range of motion of tracer particles than the increase in stiffness they induce. We attribute the difference between our result and those reported previously to the much smaller size of our myosin bipolar filaments. Nonetheless, we do observe that some of the particles in the active gels experience relatively large displacements that we attribute to motor activity, e.g. the particle trajectory in the inset of Fig.~\ref{fig:traj}b. Such large displacement are expected in active actomyosin networks (see, for example \cite{Mizuno2007,Silva2014,Toyota2011}). 

\begin{figure}
\centering
{\includegraphics[scale=0.30]{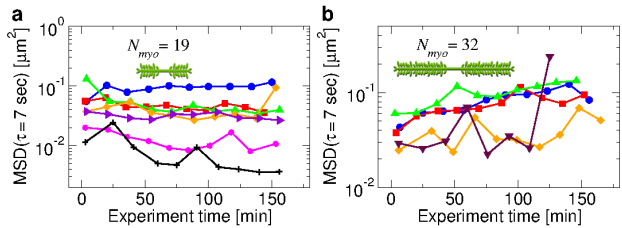}}
\caption{ MSD at $\tau=7$~s along the experiment time for actin-myosin networks constructed with myosin filaments of \textbf{a} $N_{\text{myo}}=19$ and \textbf{b} $N_{\text{myo}}=32$. Colors and symbols correspond to different [Myosin]/[Actin] ratios: 0 (blue circles), 0.0017 (red squares), 0.0025 (green triangles), 0.005 (orange diamonds), 0.0083 (violet right triangles), 0.01 (maroon down triangles), 0.0125 (magenta stars) and 0.02 (black pluses). The actin monomer concentration was 24 $\mu$M for all networks presented here. }
\label{fig:steady-state}
\end{figure}

In order to verify that our gels arrive at an active steady state we extract the time and ensemble mean square displacement (MSD) of the tracer particles. We repeat this measurement every 15 min for 160 min and compare the value of the MSD at a lag time of $\tau=7$~s. We observe (Fig.~\ref{fig:steady-state}a) that the value of the MSD of a tracer particle at this lag time changed, slightly, in the first 50 min, for most myosin concentrations, after which it maintains an approximately constant value for almost two hours. A stiffening process, therefore, takes place in the networks in the first 50 min from their polymerization, after which they remain in a steady state. For the networks with larger myosin bipolar filaments a very noisy steady state is observed (Fig.~\ref{fig:steady-state}b). We compare the time and ensemble averaged MSD of gels in the steady state and find that they coincide, indicating that these are ergodic systems (see Fig.~S5).   

\subsection{Mechanical stiffening due to myosin activity}

\begin{figure}[h!]
\centering
{\includegraphics[scale=0.30]{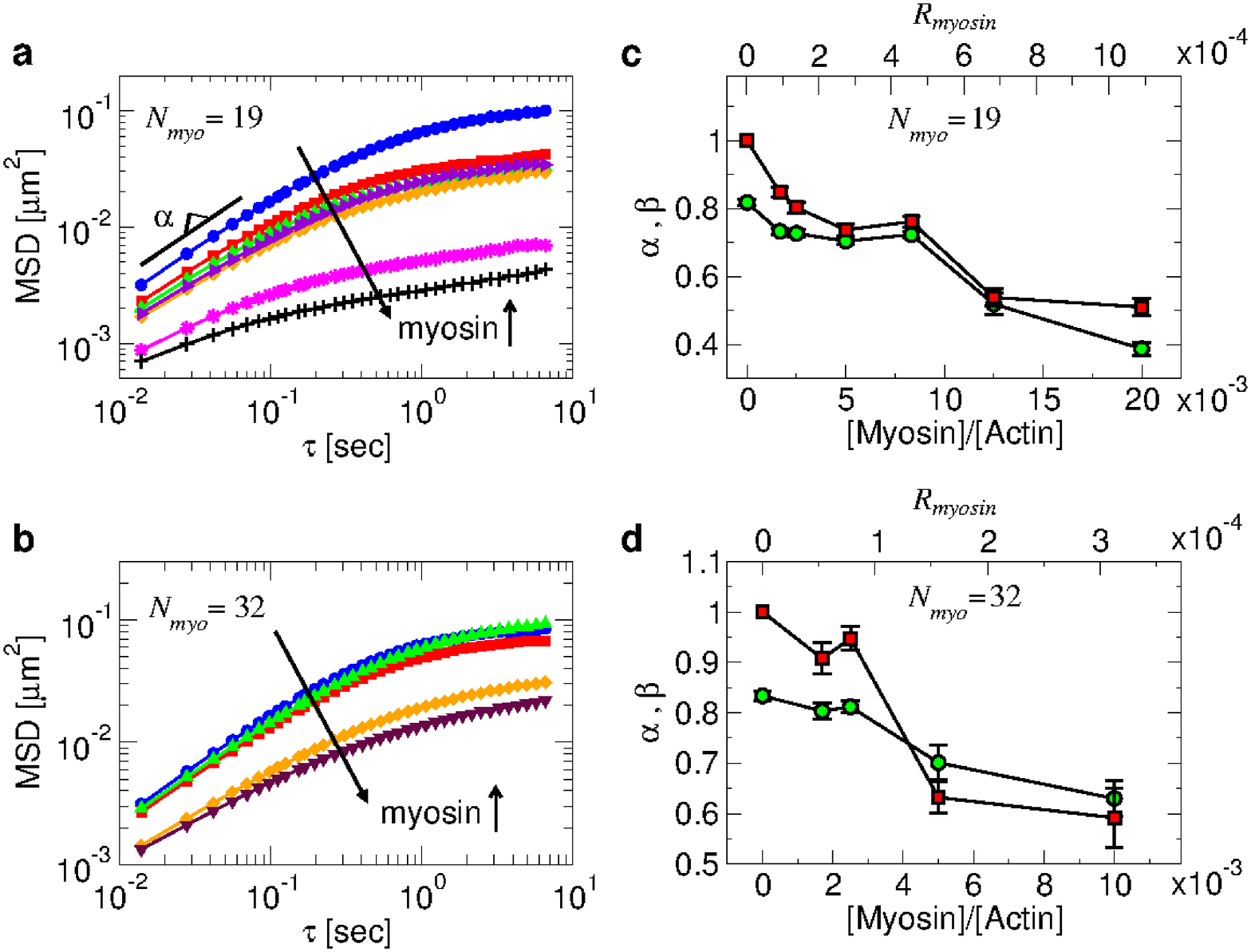}}
\caption{ Mechanical properties as a function of myosin concentration. \textbf{a,b}, Time and ensemble-averaged MSD of probe particles as a function of lag-time $\tau$ approximately 100 min after polymerization. Myosin filaments are constructed from  $N_{\text{myo}}=19$ and $N_{\text{myo}}=32$ (respectively) myosin molecules. Colors and symbols correspond to different [Myosin]/[Actin] ratios: 0 (blue circles), 0.0017 (red squares), 0.0025 (green triangles), 0.005 (orange diamonds), 0.0083 (violet right triangles), 0.0125 (magenta stars) and 0.02 (black pluses).   \textbf{c,d} stiffening of the gels as a function of myosin concentration represented by the short time diffusion exponent $\alpha$ (green circles) and the narrowing of the van Hove distribution indicated by $\beta$ (red squares) for $N_{\text{myo}}=19$ and $N_{\text{myo}}=32$ (respectively) myosin molecules (see text for details). The crosslinks concentration is given by two contributions: $R_{\text{ba}}$ and $R_{\text{myosin}}=NF_{\text{myosin}}/[\text{Actin}]$, the latter increasing with myosin concentration.}
\label{fig:mechanics}
\end{figure}

\begin{figure}[h!]
\centering
{\includegraphics[scale=0.25]{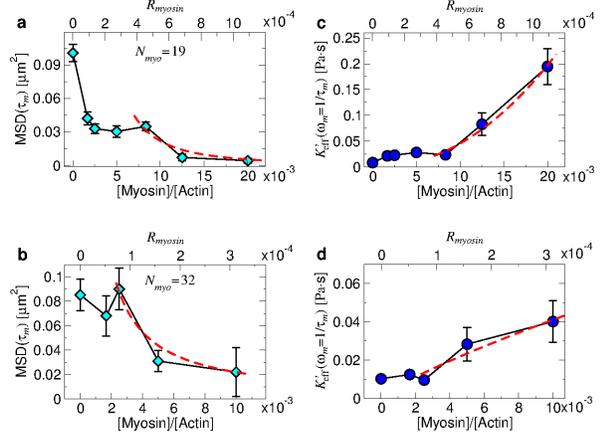}}
\caption{ \textbf{a,b} MSD at $\tau_{m}= 7$~s as a function of [Myosin]/[Actin] for $N_{\text{myo}}=19$ and $N_{\text{myo}}=32$ (respectively) reflects gel stiffening. \textbf{c,d} The corresponding lower bound effective plateau shear modulus $K'_{\text{eff}}$ as a function of [Myosin]/[Actin] for $N_{\text{myo}}=19$ and $N_{\text{myo}}=32$. Nonlinear stiffening is observe after a concentration threshold. A power law fit to the non-linear stiffening is indicated in dashed red lines. }
\label{fig:shear}
\end{figure}

The local mechanical properties of the active gels in steady state were characterized using microrheology. The MSD of tracer particles within the gels at increasing concentrations of myosin are shown in Fig.~\ref{fig:mechanics}a,b. The decrease of the MSD as myosin concentration increases in the ensemble average confirms the myosin stiffening effect observed at the single trajectory level. Since our active networks are not in thermal equilibrium, we cannot use the generalized Stokes-Einstein relation to extract their shear modulus, except for the passive network containing no myosin. Therefore, we quantify the  networks' stiffness using several measures: $\alpha$ - the diffusive exponent of the MSD curve at short times (marked in Fig.~\ref{fig:mechanics}a), $\beta$ - the ratio of active to passive width of the van Hove distribution, where the van Hove distribution is the probability density of the displacement of a tracer particle (see Fig.~S6 ), the MSD at $\tau=7$~s (lag-time within the elastic plateau of the gels), and $K'_{\text{eff}}$ - a lower bound effective differential plateau shear modulus \cite{Gardel2004} calculated using the generalized Stokes-Einstein relation.  $K'_{\text{eff}}$ is defined as the response of a prestressed material to an additional small perturbation. In Fig.~\ref{fig:mechanics}c,d $\alpha$ and $\beta$ are plotted as a function of myosin concentration. At small myosin concentrations $\alpha\approx 0.7-0.8$ in accord with the expectations for networks of semiflexible polymers \cite{Gittes1998,Morse1998,Mizuno2007}, whereas at large myosin concentrations  $\alpha\approx 0.4-0.6$ consistent with results for filaments under tension \cite{Caspi1998}. This is consisted with motor induced stiffening reported previously  \cite{Koenderink2009,Mizuno2007}. Interestingly, for networks containing large myosin bipolar filaments, that evolve dynamically and structurally with time, $\alpha$ starts at much larger values than for passive networks and decays to the passive network value \cite{Stuhrmann2012}, in contrast to our observations. 
Additional verification that the networks stiffen with the addition of myosin can be seen from the values of $\beta$ in Fig.~\ref{fig:mechanics}c,d. For all myosin concentrations $\beta<1$, indicating that the length scale of active fluctuations is small enough to be hidden in the stiffness induced narrowing of the van Hove distribution. This is in contrast to previous measurements \cite{Stuhrmann2012,Silva2014} in which $\beta>1$ was reported. This discrepancy is probably due to the smaller size of the myosin filaments in our experiments, which restricts the maximum local tension they can exert and the resulting fluctuation caused by its release. Encouraged by the small effect of active fluctuations on the shape of the van Hove distribution (see Fig.~S6), we examine the effective differential shear modulus $K'_{\text{eff}}$ of the networks as a function myosin concentrations (Fig.~\ref{fig:shear}). Since internal stresses and external stresses affect actomyosin networks equivalently \cite{Koenderink2009}, microrheology measurements in active, internally stressed networks, provide us with a measure of the differential shear modulus $K$ rather than the absolute shear modulus $G$. We find a jump in the effective differential shear modulus once myosin is added to the networks. As more myosin is added the differential shear modulus remains constant up to a threshold concentration, above which it increases again. This result is similar to the bulk rheology measurements of actomyosin networks with large myosin filaments\cite{Koenderink2009}. As mentioned above, myosin motors affect the stiffness of actin networks by crosslinking actin filaments and by pulling the slack out of the actin fibers, which reduces the networks' entropy. The plateau shear modulus of a semiflexible polymer network is given by\cite{Gittes1998,Gardel2004}:
\begin{equation}
G_0\sim\frac{\kappa^2}{k_BT\xi^2\ell_c^3},
\label{Eq:G}
\end{equation}
where  $\kappa$ is the bending modulus of an actin filament, $k_BT$ thermal energy, $\xi$ the network's mesh size, and $\ell_c$ its entanglement length (or the distance between crosslinkers in the network). The ratio between biotin/avidin crosslinks ($R_{\text{ba}}=[\text{Biotin}]/(2[\text{Actin}])$) to myosin crosslinks ($R_{\text{myosin}}=NF_{\text{myosin}}/[\text{Actin}]$) ranges, in our experiments, from $R_{ba}/R_{\text{myosin}}=8.64$ to $R_{ba}/R_{\text{myosin}}=0.43$, where $R_{\text{ba}}$ ($R_{\text{myosin}}$) is the concentration ratio between biotin/avidin crosslinks  (myosin bipolar filaments) and actin monomers. An increase in myosin concentration, therefore, increases the number of crosslinks, resulting in a decrease of the entanglement length according to $\ell_c\sim\xi R_{\text{myosin}}^{-y}$ \cite{Gardel2004}.  For actomyosin networks it was shown\cite{Koenderink2009} that $K\sim\sigma/\sigma_c$ above the critical stress $\sigma_c$. Combining this observation with Eq.~\ref{Eq:G} and assuming that internal stress increases linearly with $R_{\text{myosin}}$ we get: 
\begin{equation}
K\sim 
\begin{cases}
G_0\sim R_{\text{myosin}}^{3y},& \sigma < \sigma_c\\
G_0\frac{\sigma_c}{\sigma_0}\sim R_{\text{myosin}}^{1+3y}& \sigma \geq \sigma_c
\end{cases}
\
\label{Eq:GR}
\end{equation}
A fit of $K'_{\text{eff}}$ at high myosin concentrations to a power law resulted in $y=0.33$ and $y=0.033$ for small ($N_{\text{myo}}=19$) and large ($N_{\text{myo}}=32$) myosin filaments, respectively (Fig.~\ref{fig:shear}). For passive actin networks crosslinked with Scruin-calmodulin\cite{Gardel2004} the value of $y=0.58$ was found, similar to the power law found for our smaller motor filaments. Other reports on stiffening due to crosslinking do not differentiate between the effect of bundling and crosslinking, but find that $G_0\sim R^x$ with $x=0.6 \text{ and } 1.2$\cite{Wagner2006}, $x=0.17$\cite{Tempel1996}, and $x=1.24$\cite{Tseng2001}. It is assumed here, that the majority of the myosin filaments in the samples are attached to more than one actin filament acting as crosslinks, even though some may detach from the network occasionally. 

The onset of stiffening of the networks happens at different [Myosin]/[Actin] ratios for the two motor filaments sizes (compare 0.01 and 0.004, for $N_{\text{myo}}=19$ and $N_{\text{myo}}=32$, respectively), and at different $R_{\text{myosin}} $ (compare $5.3\cdot 10^{-4}$ and $1.2\cdot 10^{-4}$, for $N_{\text{myo}}=19$ and $N_{\text{myo}}=32$, respectively). However, both are in a similar range to previous reports\cite{Koenderink2009,Bendix2008}.
\subsection{Stress propagation in active networks}

Two point microrheology is a technique that uses the spatial and temporal dependence of the correlated diffusion of particle pairs to measure the bulk shear modulus of the medium in which the particles are embedded \cite{Crocker2000}. In addition, the correlated diffusion can be used to measure the spatial decay of the material's response to a mechanical perturbation \cite{Diamant2015,Oppenheimer2011,Cui2004}. The correlated diffusion in such experiments is calculated from particle trajectories according to\cite{Crocker2000}:
\begin{eqnarray}
D_\parallel(r,\tau) &=& \langle\Delta r^i_\parallel(t,\tau) \Delta r^j_\parallel(t,\tau)
\delta(r-R^{ij}(t))\rangle \nonumber\\
D_\perp(r,\tau) &=& \langle\Delta r^i_\perp(t,\tau) \Delta r^j_\perp(t,\tau)
\delta(r-R^{ij}(t))\rangle,
\label{drr}
\end{eqnarray}
where $\Delta r^i_\parallel(t,\tau)$ ($\Delta r^i_\perp(t,\tau)$) is
the displacement of particle $i$ during the time between $t$ and
$t+\tau$, projected parallel (perpendicular) to the line connecting
the pair, and $R^{ij}(t)$ is the pair separation at time $t$. The shear modulus of the material is then extracted from the Laplace transform of the longitudinal part of the correlated diffusion using the relation  $\tilde{D}_\parallel(r,s)=k_BT/6\pi r s \tilde{G}(s)$, where $\tilde{G}(s)=s\tilde{\eta_{b}}(s)$, $s=i\omega$, and $\tilde{\eta_{b}}(\omega)$ is the complex bulk complex viscosity. 

Recently, we showed that in passive actin networks (without myosin) such a perturbation, over an intermediate range of distances, decays much faster with distance ($r^{-3}$) than is usually expected ($r^{-1}$) \cite{Sonn2014,Sonn2014b,Sonn-Segev2017}. This response is a manifestation of the transition in the motion of the solvent with respect to the polymer network. We assume no-slip boundary condition between the solvent and the particle surface, but the actin network can slide over it \cite{Levine2000}. As a result, in response to a mechanical perturbation, the fluid moves against the actin network  at intermediate distances, whereas  the solvent and network move together at long distances. The transition between these two scenarios occurs at particle separations of several micrometers in actin networks. The crossover distance, $r_c$, depends on the actin mesh size and the ratio between the bulk and local complex viscosity, $\eta_b$ and $\eta_\ell$ respectively, as\cite{Sonn2014,Diamant2015}:
\begin{equation}
r_c = a [2(\eta_b/\eta_\ell)g(\xi_d/a)]^{1/2},\ \
g(x) = x^2+x+1/3.
\label{rc}
\end{equation}
where $a$ is the tracer particle radius, the local complex viscosity is the complex viscosity extracted from one point microrheology, while the bulk complex viscosity is extracted from two point microrheology measurements. Eq.~\ref{rc} provides a means to extract structural information, i.e. the dynamic correlation length $\xi_d$, about the networks, from microrheology experiments \cite{Sonn2014b}. In actin networks made of sufficiently long filaments $\xi_d=b\xi$, where $b$ is a constant close to one\cite{Sonn2014b}. 

\begin{figure}
\centering
{\includegraphics[scale=0.27]{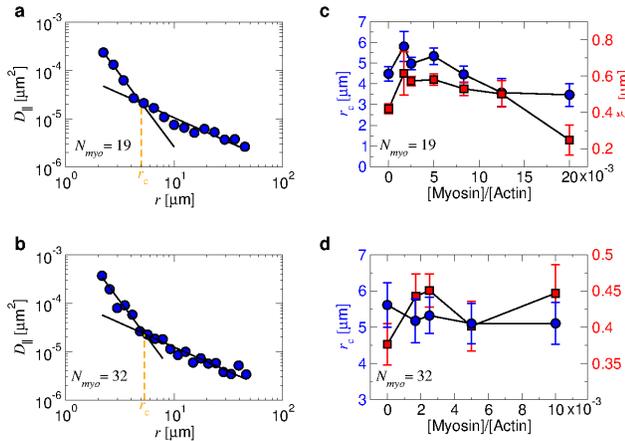}}
\caption{Longitudinal displacement correlations as a function of particle separation at lag
	time $\tau=0.014$~s at [Myosin]/[Actin]=0.0025 with  $N_{\text{myo}}=19$ (\textbf{a}) and  $N_{\text{myo}}=32$ (\textbf{b}). The errors of $D_{||}$ are approximately $5\cdot10^{-6}~\mu$m$^2$ or smaller. The crossover
	distance (orange dashed line) is defined at the intersection of the fitted bulk ($r^{-1}$) and intermediate ($r^{-3}$) power-laws decays
	of $D_\parallel$.  The cross over distance $r_c$ and the dynamic correlation length $\xi_d$ dependence on [Myosin]/[Actin] for $N_{\text{myo}}=19$ (\textbf{c}) and $N_{\text{myo}}=32$ (\textbf{d}).}
\label{fig:2p}
\end{figure}

Here, we are interested in characterizing stress propagation in active networks as a function of motor concentration. The increased stiffness due to motor action translates to an increase in bulk complex viscosity, and hence, an increase in $r_c$. On the other hand, an increase in motor concentration and crosslinks should result in a decrease in mesh size and a decrease in $r_c$. In Fig.~\ref{fig:2p}a,b the longitudinal correlated diffusion at $\tau=14$~ms is shown, which resembles the measured $D_\parallel$ of passive actin gels with no myosin motors \cite{Sonn2014,Sonn2014b}.  Specifically, the correlated diffusion decays fast, $D_\parallel\sim r^{-3}$, at intermediate distances and slower, $D_\parallel\sim r^{-1}$, at large distances, as in passive actin networks. Based on the small symmetric effect that active fluctuations have on the van Hove distribution (Fig.~\ref{fig:mechanics}c,d and Fig.~S6), we assume that Eq.~\ref{rc} is valid also for the active networks at steady state. We use the effective lower bound complex viscosity measured by one and two point microrheology, and the crossover distance $r_c$, to extract the dynamic correlation length $\xi_d$ as a function of myosin concentration (Fig.~\ref{fig:2p}c,d). The crossover distance $r_c$, ranges in our active systems between $4.5-3.5$ $\mu$m and  $5.5-5.0$ $\mu$m for $N_{\text{myo}}=19$ and $N_{\text{myo}}=32$ motors respectively. In both motor systems we find that the dynamic correlation length increases with a small addition of myosin motor (Fig.~\ref{fig:2p}c,d). For the smaller motor filaments, we find that increasing further the myosin concentration causes a  decreases in $\xi_d$.  For larger motor filaments, we find that $\xi_d$ is larger than its value in the passive network and approximately constant at different myosin concentrations, however, $r_c$ reduces with the addition of myosin. We conclude, therefore, that while motor activity increases the networks stiffness and may change its structure (i.e., mesh size), it has little effect on the spatial dependence of stress transmission, since changes in $r_c$ are minor.  

\section{Conclusions}
In this paper we presented a bio-inspired gel that can exist in a long-lived active steady state. This allowed us to study the effect of activity level, in terms of motor size and concentration, on the mechanical properties of these active networks. Steady state conditions were obtain approximately 50 min after actin polymerization was initiated, probably after some initial restructuring of the network. Using microrheology, we have shown that myosin motors stiffen in-vitro actin networks, probably by the addition of crosslinking points and by applying tension to the network. The crosslinking effect is quantified by measuring the change is the networks' mesh size using two point microrheology as well as by the power law increase of the effective shear modulus. The tension effect is seen through the decrease in diffusion power at short times, $\alpha$. These two stiffening mechanisms where found also in networks containing much larger myosin bipolar filaments and network in which crosslinkers induce actin bundling. The similarity between the effects of motor activity on the mechanical properties of structurally different actin networks, indicates that this type of stiffening should apply for the more general case of cytoskeleton networks. We note that the kinetics of self-organization of cytoskeletal networks should differ significantly from these in-vitro networks, for several reasons. First, the binding kinetics of nonmuscle myosin present in most cells are different than the kinetics of the rabbit skeletal myosin II we use. Second, there are more than 50 actin binding proteins in the cells that take an active role in its assembly, and are not present in our model system. Finally, actin turnover in cells is known to affect self-organization, as seen in cell motility. Actin turnover in cells may also act as a stress regulation mechanism. For example, there is evidence of such interplay between mechanical and biochemical processes in actin stress fibers in the cell \cite{Stachowiak2014}. However, working with small myosin bipolar filaments, in vitro, allowed us to avoid the catastrophic rupture and compression usually seen in such networks, suggesting the motor bipolar filaments size as a possible additional control parameter for structural stability within cells.  

In addition to the relevance that these systems may have to biological systems, they serve also as model active materials. The long lived steady state demonstrated here provides a much needed opportunity to study, experimentally, statistical mechanics properties of active systems coupled to a heat bath. For example, we observed a distinct difference between tracer particle fluctuation in our gels as compared to previously studied actomyosin networks containing large motor filaments\cite{Stuhrmann2012,Silva2014,Mizuno2007,Toyota2011}. Relatively small filaments cause small athermal fluctuations leading to strongly subdiffusive MSD curves (i.e. with a diffusion exponent $\alpha<0.7$), whereas large myosin filaments induce large, non-Gaussian, fluctuations with $\alpha\sim 0.9$ \cite{Stuhrmann2012,Silva2014}. 
Another interesting question that could be addressed in these systems is the existence of detailed balance. The probability density currents of particle motion (see for example\cite{Gladrow2016}) in this active steady state, might differ at different timescales, especially if network reorganization and evolution takes place.

\section*{Acknowledgments}

The authors are grateful to Haim Diamant for numerous illuminating discussions. This research was supported by the Marie Curie Reintegration Grant (PIRG04-GA-2008-239378), the Israel Science Foundation grant 1271/08, and by the US-Israel Binational Science Foundation grant 2014314. A. S.-S acknowledges funding from the Tel-Aviv University Center for Nanoscience and Nanotechnology. A. B.-G. acknowledges funding from the Israel Science Foundation (grant 1618/15). 

%


\end{document}